\documentclass[%
 reprint,
 amsmath,amssymb,
 aps,
 prb,
floatfix,
]{revtex4-2}

\usepackage{graphicx}% Include figure files
\usepackage{dcolumn}% Align table columns on decimal point
\usepackage{bm}% bold math
\usepackage[ruled,vlined]{algorithm2e}
\usepackage{calc}

\SetKwFunction{FChooseW}{ChooseWAssignments}
\SetKwFunction{FChooseB}{ChooseBAssignments}
\SetFuncSty{textsc}

\usepackage{multirow}
\usepackage{array}
\usepackage{booktabs}
\usepackage{siunitx}
\usepackage{xspace}
\usepackage[breaklinks=true]{hyperref}% add hypertext capabilities
\usepackage[capitalize, noabbrev]{cleveref}
\crefname{algocf}{algorithm}{algorithms}
\Crefname{algocf}{Algorithm}{Algorithms}
\usepackage{orcidlink}
\usepackage{placeins}
\newcommand{\pairton}{Pairton\xspace}

\mathchardef\mhyphen="2D
\newcommand{\lrp}[1]{\left( #1 \right)}

\newcommand{\lrc}[1]{\left\lbrace #1 \right\rbrace}
\newcommand{\abs}[1]{\left \vert #1 \right \vert }

\newcommand{\EE}{\mathbb{E}}
\newcommand{\PP}{\text{\usefont{U}{bbm}{m}{n}P}}

\DeclareSIUnit{\electronvolt}{e\kern-0.1em\relax{}V}
\DeclareSIUnit{\eV}{\electronvolt}

\DeclareMathOperator{\logit}{logit}

\begin{document}

% \preprint{APS/123-QED}

\title{\pairton: Iterative Reconstruction of Short-Lived Particles}% Force line breaks with \\

\author{Andreas Hermansen\orcidlink{0009-0006-1162-9770}}
\email{andreas.hermansen@unige.ch}
\author{Chris Scheulen\orcidlink{0000-0002-9142-1948}}
\email{chris.scheulen@cern.ch}
\author{Tobias Golling\orcidlink{0000-0001-8535-6687}}
\affiliation{%
    Département de physique nucléaire et corpusculaire, University of Geneva
}

\date{\today}% It is always \today, today,

\begin{abstract}
    We present \pairton, an iterative framework for reconstructing short-lived particles in high-energy collision events.
    By formulating particle reconstruction as a masked prediction process over graph structures, \pairton learns conditional distributions consistent with a factorised decomposition of decay products and iteratively predicts edges in the adjacency matrix representing particle decay relationships.
    Leveraging a pairformer-based architecture with dynamically updated pairwise representations, our method incorporates global event consistency.
    We demonstrate state-of-the-art performance on fully hadronic \(t\bar{t}\) decays.
    \pairton provides a general, flexible paradigm for particle reconstruction and can be readily extended to other topologies, bridging ideas from modern generative modelling and high-energy physics.
\end{abstract}

\maketitle

%\tableofcontents

\section{\label{sec:intro} Introduction}
At the Large Hadron Collider (LHC)~\cite{lhc_standard}, short-lived particles, such as top quarks and Higgs bosons, are produced in abundance by colliding beams of protons at unprecedented energies.
These particles decay almost instantaneously to form leptons and hadronic jets, collimated showers of hadrons, that can be detected by the various sub-detectors surrounding the collision point. 
Reconstructing these short-lived particles from their decay products is a crucial step in studying their properties and interactions.

In processes with high particle multiplicities, such as those involving fully hadronic decay of a top quark pair, the combinatorial background can be significant.
As a result, it is challenging to accurately identify the decay products of the short-lived particles.

In this paper, we propose \pairton, a generative perspective on particle reconstruction.
We formulate the problem as structured probabilistic inference over partially observed graphs, where nodes represent final state objects and edges encode their decay relationships.
\pairton reconstructs missing edges iteratively by predicting them conditioned on previously inferred structure.

We apply \pairton to the common benchmark of reconstructing the all-hadronic decay mode of top quark pairs (\(t\bar{t}\)), where the full decay chain is
\begin{equation*}
    t\bar{t} \to W^{+}b\,W^{-}\bar{b} \to q_1\bar{q}_1'b\,q_2\bar{q}_2'\bar{b}.
\end{equation*}
All six quarks hadronise and can be detected as jets in the detector, whereby the two jets originating from \(b\) quarks are referred to as \(b\)-jets, which can be distinguished from the jets originating from lighter quarks. 
To fully reconstruct the \(t\bar{t}\) system, one therefore needs to correctly identify which jets originate from the \(W\) bosons and which jets originate from the \(b\) quarks.
Additional jets from gluon radiation and pile-up interactions are possible in the collision events and complicate this task. 
Methods developed and tested on the \(t\bar{t}\) system can additionally be extended to other final states involving short-lived particles.

The implementation of \pairton and the workflows for training, inference, and ablation studies are available at:
\texttt{\href{https://github.com/AMHermansen/pairton}{https://github.com/AMHermansen/pairton}}.

\section{\label{sec:current_approaches} Current Approaches}
Traditional methods for short-lived particle reconstruction, such as the likelihood-based KLFitter algorithm~\cite{klfitter}, often rely on a combinatorial approach, where all possible combinations of detected particles are considered as potential decay products.
In these algorithms, the combination which achieves the lowest value of a predefined cost function is selected as the best candidate.
Alternatively, multiple machine learning methods have been developed in recent years to address this task:

SPANet~\cite{spanet1, spanet2, spanet3, spanet4} uses a transformer-based architecture with an attention mechanism to capture the underlying symmetries of the system.

Topograph~\cite{topograph} views the problem as a graph reconstruction task, wherein top quark and \(W\) boson candidates are injected as additional nodes, and the model classifies edges between jets and these candidates.
The idea of framing the problem as a graph reconstruction task without the need for additional top and \(W\) boson nodes has already been discussed by the Topograph authors, but was not explored further there.

HyPER~\cite{hyper} uses a hypergraph representation, where hyperedges of degree two represent the decays of \(W\) bosons, and hyperedges of degree three represent the decays of top quarks.

TIGER~\cite{tiger} creates a topology agnostic, hierarchical graph neural network to infer particle decay assignments without assuming a fixed event topology.
As a result, the model allows for general-purpose reconstruction and classification across varied processes.

All of these machine learning methods achieve enhanced performance compared to traditional methods~\cite{klfitter, spanet1, topograph, hyper, tiger}.

\section{\label{sec:method} Method}

The reconstruction of the all-hadronic \(t\bar{t}\) topology involves assigning the correct final-state objects of two \mbox{\(t \to Wb\)} decays when neglecting the differentiation into particles and anti-particles.
As a result, one has to find the two jets belonging to the \(W\) boson produced in each top quark decay, and assign a third jet for the \(b\) quark produced in the same decay.

Generally, the precise assignment of the jets initiated by the quark and the anti-quark in the \(W\) boson decay is not performed in most physics analyses, such that permutations among these are permitted.
Similarly, permuting the top quark and the anti-top quark is equally allowed.
In summary, we are therefore interested in assigning a pair of jets for each \(W\) boson decay, and a third jet belonging to the corresponding \(b\) quark, resulting in a total of six assigned jets.

Mathematically, the optimal reconstruction can be expressed as the most probable reconstruction,
\begin{equation}
    \hat{W}_1, \hat{W}_2, \hat{b}_1, \hat{b}_2= \arg \max_{W_1, W_2, b_1, b_2} \PP\lrp{W_1, W_2, b_1, b_2 \mid x},
\end{equation}
where \(x\) represents the data associated with the event, and \(W_1\), \(W_2\), \(b_1\), \(b_2\) represent the reconstructed \(W\) bosons and \(b\)-jets, respectively.
We note that, while \(b_i\) represents a single index, \(W_i\) represents an unordered set of two indices, when assigning a unique index to each jet in the collision event under consideration.

By the chain rule of probability, the joint distribution factorises as the product of conditional probabilities.
In the case of reconstructing an all-hadronic \(t\bar{t}\) decay, this results in
\begin{equation}
    \begin{aligned}
        &\PP\lrp{W_1, W_2, b_1, b_2 \mid x} \\
        &= \PP(W_1 \mid x) \\
        &\hphantom{=}\:\times \PP(W_2 \mid W_1, x) \\
        &\hphantom{=}\:\times \PP(b_1, b_2 \mid W_1, W_2, x),
    \end{aligned}
    \label{eq:factorization}
\end{equation}
where the indexing of the \(W\) bosons and \(b\) quarks is arbitrary.
Each of the conditional probability terms can be understood as a partial event reconstruction step of the corresponding particle if the reconstruction is performed iteratively.

Ablation experiments included in \cref{sec:appendix_model_ablation} show no measurable benefit from further factorising the distribution of the \(b\)-jets.
Therefore, we keep the \(b\)-jets together in the last term to reduce the number of function evaluations on our model.
For completeness, we list the validation performance of fully factorising the joint distribution in \cref{sec:appendix_model_ablation}.

To learn the conditional distributions, we reframe the problem using a graph representation of the particle assignments in order to deploy a masked denoising algorithm in our approach.
As a result, the conditional probabilities introduced in \cref{eq:factorization} can naturally be represented in iteratively updated graphs, which we feed back into the model in subsequent particle assignments.
In \cref{sec:appendix_proof}, we establish the formal connection between the factorised probability distribution and our masked denoising procedure, which we present in the following.

\subsection*{Graph Representation}

\begin{figure}[htb]
    \centering
    \includegraphics[width=0.48\textwidth]{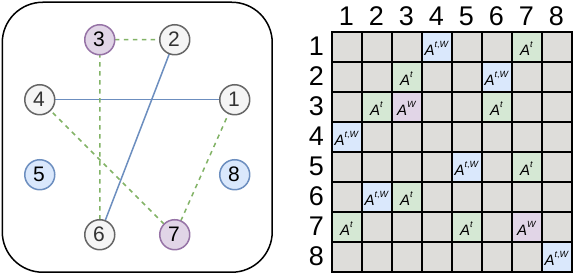}
    \caption{%
        Representation of an all-hadronic \(t\bar{t}\) event with eight jets as (left) a graph of the decay products of the individual tops and (right) the combined view of the corresponding top and \(W\) boson adjacency matrices \(A^t\) and \(A^W\), respectively.
        In the chosen example, one top quark decays into the jets 1, 4, and 7, where jet 7 originates from the \(b\) quark while jets 1 and 4 originate from the subsequent \(W\) boson decay.
        Similarly, the other top quark decays into jets 2, 3, and 6 with jet 3 originating from the \(b\) quark.
        Jets 5 and 8 are additional jets not originating from the top quark decays.
        In the adjacency matrix view, grey entries correspond to zeros, while the coloured entries correspond to graph edges (including self-loops) corresponding to the graph assignment on the left:
        Blue entries and solid edges are ones in both \(A^t\) and \(A^W\), green entries and dashed edges are ones in \(A^t\) only, and purple entries are ones in \(A^W\) only.
        \label{fig:graph_representation}
    }
\end{figure}

We represent each event as a collection of simple graphs \mbox{\(\mathcal{G}_\alpha = (\mathcal{V}, \mathcal{E}_\alpha)\)} with a shared set of nodes. Each graph corresponds to a particle type \mbox{\(\alpha \in \{\mathrm{W}, t\}\)} for the all-hadronic \(t\bar{t}\) decays that we are considering.
The node set \(\mathcal{V}\) corresponds to detected jets.
An edge \mbox{\((i,j) \in \mathcal{E}_\alpha\)} indicates that jets \(i\) and \(j\) are decay products of particle \(\alpha\).
Self-loops encode that a jet is not associated with the corresponding particle type.

This representation allows us to encode particle assignments as adjacency matrices \mbox{\(A^\alpha = {a_{ij}^\alpha}\)} defined through
\begin{equation}
    a_{ij}^\alpha = \begin{cases}
        1 \quad &\text{if } \lrp{i,j} \in \mathcal{E}_\alpha, \\
        0 \quad &\text{otherwise}.
    \end{cases}
\end{equation}
This is illustrated in \cref{fig:graph_representation}.

Beyond providing a natural representation of the physical decay topology, formulating the problem via adjacency matrices offers an architectural advantage.
It allows the partially observed adjacency matrices, \(\tilde{A}^\alpha\), to be directly embedded and incorporated into the model as dynamic pairwise edge features.
By treating these structural assignments as input features, a graph neural network can seamlessly condition its subsequent predictions on the explicitly represented relational state between all jets without requiring complex message-passing operations over higher-order tensors.

\subsection*{Masked Denoising Formulation}

Inspired by discrete diffusion~\cite{d3pm} we formulate the reconstruction problem as a discrete denoising problem over adjacency matrices defined above.
Let \(A\) denote the ground-truth adjacency matrices.
We then define a partially-deterministic Markov chain responsible for the forward masking process, which produces partially observed adjacency matrices \(\tilde{A}\),
\begin{equation}
    q_t(\tilde{A} \mid A), \quad t \in \{0,1,2\},
\end{equation}
where the masking level \(t\) determines which particles are visible:
\begin{itemize}
    \item both \(W\) bosons revealed,
    \item one \(W\) boson revealed,
    \item no particles revealed.
\end{itemize}
Entries logically inferable from revealed particles are filled with their true values in \(\tilde{A}\), while remaining entries are replaced with a mask token \(\mathrm{M}\).
This is a special case of the absorbing state discrete diffusion processes used in language modelling adapted to the physical system.
Instead of relying on fully random masking directly applied to the adjacency matrices, we opt for this particular Markov chain to accommodate the hierarchical nature of the \(t\bar{t}\) decay, but note that any Markov chain that eventually ends up with no revealed particles will work.
This particular Markov chain effectively exploits the physical nature of the \(t\bar{t}\) decay to define an efficient corruption scheme allowing us to reverse the process using few steps.

The hierarchical corruption scheme directly mirrors the physical decay topology of the \(t\bar{t}\) system, allowing the model to unambiguously condition on previously predicted particles at each step of the reverse process.
Conversely, an unstructured approach that randomly masks individual edges instead would frequently yield intermediate configurations lacking a clear physical interpretation.
Furthermore, such random masking risks creating states where missing edges could be deterministically inferred due to the strict structural constraints of the adjacency matrices.
The simplicity of the above process is a consequence of the hierarchical and symmetric nature of the \(t\bar{t}\) process, which limits the number of meaningful intermediate masking states.
More complicated topologies like the associated production of a Higgs boson with a top quark pair, \(t\bar{t}H\), would have a greater number of semantically meaningful masking states, such as one visible \(W\) boson or one visible Higgs boson, which in turn would lead to richer and more expressive Markov processes.

\begin{figure}[tb]
    \centering
    \includegraphics[width=0.48\textwidth]{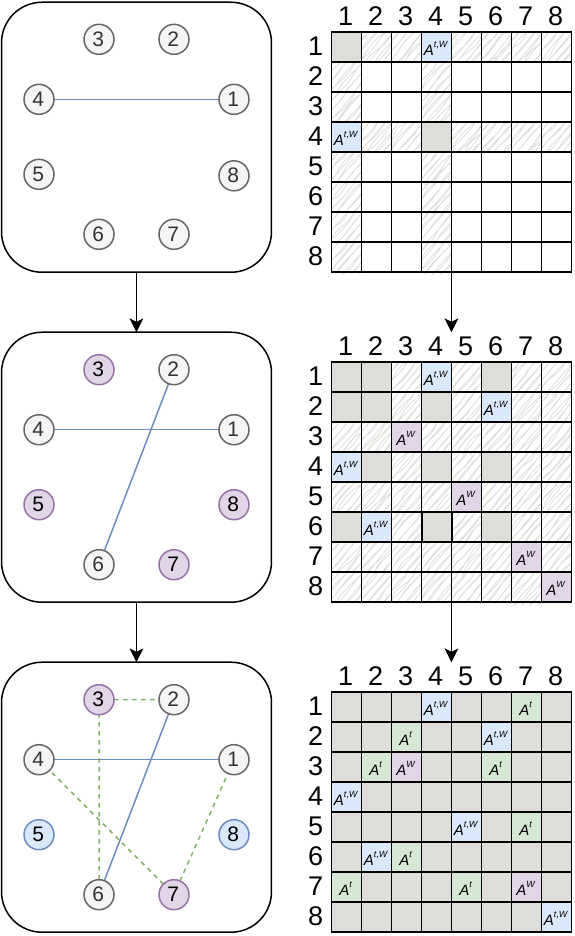}
    \caption{%
        Sequential reconstruction of \(W\) bosons and top quarks in three steps for the event depicted in \cref{fig:graph_representation} following the same visual style.
        Additionally, white entries indicate unknown adjacency, while the scratched entries indicate that no edge can exist in the \(W\) boson edge-set, so that the associated jets cannot be decay products of the same \(W\) boson, but might still be part of the top quark edge-set.
        \label{fig:graph_reconstruction}
    }
\end{figure}

We train a model \mbox{\(f_\theta(\tilde{A}, x)\)} to predict masked entries conditioned on the observed structure and jet features.
The training objective in this case is
\begin{equation}
    \mathcal{L}(\theta)
        = \EE \Bigg[
        \sum_{\alpha} \sum_{i,j \in \mathcal{V}}
        \delta_{\mathrm{M}}(\tilde{a}_{ij}^\alpha) \,
        \ell_{\mathrm{bce}}\big(a_{ij}^\alpha, (f_\theta(x, \tilde{A}))_{ij}^\alpha\big)
        \Bigg],
\end{equation}
where \(\delta_{\mathrm{M}}\) is 1 if the entry is masked and 0 otherwise, and the expectation runs over both the training data and over the masking level \(t\).
This objective corresponds to masked conditional likelihood training, analogous to the discrete diffusion models introduced by LLaDA~\cite{llada} for natural language processing.
More broadly, this paradigm of absorbing state discrete diffusion is a compelling alternative generative framework for large language models compared to the more traditional nature of left-to-right auto-regressive decoding~\cite{d3pm, mdlm, radd, sedd}.

While masked and predictive methods have gained traction in high-energy physics, they have largely been used in BERT-style models with a fixed masking rate (e.\,g.\ Masked Particle Modeling on Sets~\cite{mpm1}) or standard GPT-style auto-regressive masking (e.\,g.\ OmniJet-\(\alpha\)~\cite{omnijet_alpha}). 
Our work introduces discrete diffusion to high-energy physics.
By formulating event reconstruction as a discrete diffusion process over the adjacency matrices, our method learns to reverse the masking process, which enables it to iteratively predict the structural topology of an event.

At inference time, reconstruction proceeds by iteratively predicting particle assignments, thus corresponding to the reverse of the masking process.
Starting from a fully masked adjacency matrix, the model predicts particles and refines new predictions with previously selected structures, reflecting the factorisation as outlined in \cref{eq:factorization}.
A visual illustration of this approach is provided in \cref{fig:graph_reconstruction}.

We approximate the maximum a posteriori (MAP) assignment by greedily selecting edges that maximise the sum of predicted logits,
\begin{equation}
    \sum_{i \leq j} Z_{ij} \logit p_{ij},
    \label{eq:logit_heuristic}
\end{equation}
where \(Z_{ij}\) indicates selected edges, and \mbox{\(p_{ij} = f_\theta(x, \tilde{A})_{ij}\)} is the corresponding model output. We note that it would be possible to explore more sophisticated sampling algorithms like beam searches~\cite{beam_search}, but leave the exploration of such methods for future work.
The pseudo-code for our full reconstruction algorithm is included in  \cref{sec:appendix_algorithm}.

It is important to note that while our algorithm used at inference is deterministic, our model is learning the conditional distributions outlined in \cref{eq:factorization} as shown for completeness in \cref{sec:appendix_proof}.

\subsection*{Model Architecture\label{subsec:architecture}}

To model \mbox{\(p_\theta(a_{ij} \mid \tilde{A}, x)\)}, we use a pairformer-based architecture initially developed as the primary component of the AlphaFold3 architecture for protein structure prediction~\cite{alphafold3}.
Conceptually, the architecture is a modified version of a transformer, where the model also uses pairwise features to augment the attention scores, \(S_{ij}\), with an additional bias term
\begin{equation}
    S_{ij} = \frac{q_i \cdot k_j}{\sqrt{d}} + b_{ij}.
\end{equation}
Here, \(b_{ij}\) is a learnable function of the pairwise features.
This is conceptually very similar to the particle attention mechanism used in the Particle Transformer (ParT), which has achieved remarkable performance in jet tagging~\cite{ParT}.
However, where ParT uses a fixed pairwise representation, the pairformer architecture dynamically updates the pairwise features.

\begin{figure}[tb]
    \centering
    \includegraphics[width=0.48\textwidth]{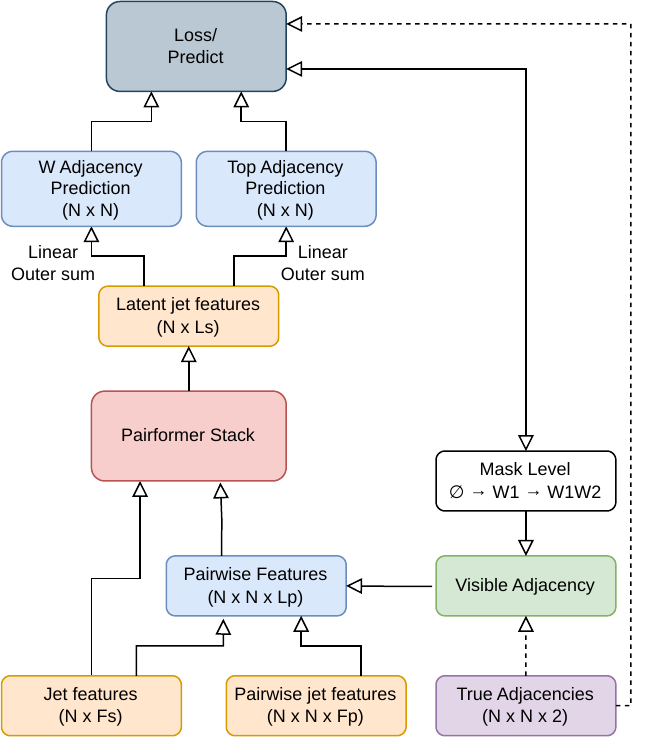}
    \caption{%
        Visual illustration of the model during training and inference.
        During training, the mask level is chosen at random and used together with the true adjacency matrices to construct the visible adjacency matrices given to the model.
        During inference, the visible adjacency matrices are constructed iteratively.
        The true adjacency matrices are also used as targets for the loss calculation during training.
        \label{fig:full_model}
    }
\end{figure}

Edge predictions are computed as
\begin{equation}
    \hat{a}_{ij}^\alpha = g_\phi^\alpha \big(W \tilde{x}_i + W \tilde{x}_j \big),
\end{equation}
where \(W\) is a learnable linear projection, \(g_\phi^\alpha\) is an MLP, and \(\tilde{x}_i\) are latent node embeddings.
A visual model representation during training and inference is shown in \cref{fig:full_model}.

\begin{table}[tb]
    \caption{%
        Feature transformation applied for both single- and pairwise jet features.
        Only features for which transformations are applied are listed, while the other features mentioned in the text are propagated to \pairton as is.
        \label{table:feature_preprocessing}
    }
    \begin{tabular}{l l}
        \toprule
            Feature\quad~                            & Feature(s) after Transformation              \\
        \midrule
            \(E\)                                    & \(\log(E / \unit{\giga\eV}) - 4.5\)          \\
            \(p_\text{T}\)                           & \(\log(p_\text{T} / \unit{\giga\eV}) - 4.5\) \\
            \multirow{2}{*}{\(\varphi\)}             & \(\cos \varphi\)                             \\
                                                     & \(\sin \varphi\)                             \\
        \midrule
            \multirow{3}{*}{\(\Delta \varphi_{ij}\)} & \(\Delta \varphi_{ij}\)                      \\
                                                     & \(\Delta (\sin \varphi)_{ij}\)               \\
                                                     & \(\Delta (\cos \varphi)_{ij}\)               \\
            \(m_{ij}\)                               & \(\log(m_{ij} / \unit{\giga\eV}) - 4.5\)     \\
        \bottomrule
    \end{tabular}
\end{table}

As the input jet features, we use the jet energy \(E\), transverse momentum \(p_\text{T}\), pseudo-rapidity \(\eta\), azimuthal angle \(\varphi\), and whether or not the jet was \(b\)-tagged.
Additionally, we use the pairwise jet features of the di-jet invariant mass \(m_{ij}\), the difference in pseudo-rapidity \(\Delta \eta_{ij}\) and azimuthal angle \(\Delta \varphi_{ij}\) between the jets, and the angular distance \mbox{\(\Delta R_{ij} = \sqrt{(\Delta \varphi_{ij})^2 + (\Delta \eta_{ij})^2}\)} of the jets from each other.
During the model pre-processing, some of these input features are transformed as listed in \cref{table:feature_preprocessing}.
Additionally, the visible adjacency matrices determined from the true adjacency matrices and the mask step during training or from the previous inference steps during inference are embedded through an embedding layer and added to the embedded pairwise jet features before passing the features into the pairformer stack as shown in \cref{fig:full_model}.

\section{\label{sec:results} Results}

\subsection*{Dataset}
For model training and evaluation, the dataset originally published in conjunction with HyPER~\cite{hyper_data} was used.
It contains a total of \num{6e7} all-hadronic \(t\bar{t}\) events simulated at a centre-of-mass energy of \mbox{\(\sqrt{s} = \qty{13}{\tera\eV}\)} using \textsc{MadGraph5}\_a\textsc{MC@Nlo}\,2.9.16~\cite{mg5} at next-to-leading order in QCD for the matrix element calculation and \textsc{Pythia}\,8.306~\cite{pythia} for the parton shower.
For the detector response, \textsc{Delphes}\,3.5.0~\cite{delphes} was used in a configuration simulating the ATLAS detector~\cite{atlas}.

Jets were reconstructed using the anti-\(k_t\) algorithm~\cite{antikt} implemented in \textsc{FastJet}~\cite{fastjet} with a radius parameter of \mbox{\(R = 0.4\)}.
The resulting jets are required to pass a transverse momentum threshold of \mbox{\(p_\text{T} = \qty{25}{\giga\eV}\)} and a cut in the absolute pseudo-rapidity of \mbox{\(\abs{\eta} < 2.5\)}.
The tagging of \(b\)-jets was simulated by applying a \(p_\text{T}\)-dependent tagging efficiency based on Reference~\cite{btag_efficiency} to jets containing \(b\)-hadrons.
Reconstructed objects were matched to final-state particles of the matrix element generator with a maximal angular distance of \mbox{\(\Delta R < 0.4\)}.
All events are required to contain at least six jets, of which at least two are \(b\)-tagged.

The dataset containing events passing the selection criteria was randomly partitioned into training, validation, and testing sub-sets containing \qty{90}{\percent}, \qty{5}{\percent}, and \qty{5}{\percent} of events, respectively.
After applying the event selection criteria described above and requiring at least one fully reconstructible \(W\) boson in each event, roughly \num{8.8e6} events remain in total for training and validating the models.
Out of these, roughly \num{2.4e6} are fully reconstructible.
For the exact numbers and a complete description of the event generation setup we refer to Reference~\cite{hyper}.

\subsection*{Reconstruction Performance and Statistical Stability}

As the central performance metric, we record the efficiencies, also referred to as true positive rates, of reconstructing full events (\(\varepsilon_{t\bar{t}}\)), top quarks (\(\varepsilon_{t}\)), and \(W\) bosons (\(\varepsilon_{W}\)), representing a common way of determining the performance of reconstruction algorithms~\cite{spanet1, topograph, hyper}.
Due to the high degree of symmetry of the fully hadronic \(t\bar{t}\) decay, we record a \(W\) boson as correctly reconstructed if the two associated jets are correctly identified, although they may be permuted.
For a top quark to be correctly reconstructed, its daughter \(W\) boson and the remaining \(b\) quark must be correctly predicted.
For the full \(t\bar{t}\) system to be correctly reconstructed, both the top and anti-top quark have to be correctly reconstructed.
However, the prediction of the top and anti-top quarks may be permuted, since jet charge measurements of the decay products of fully hadronic \(t\bar{t}\) decays are challenging and commonly not attempted in analyses of the process.

\begin{table}[tb]
    \centering
    \caption{%
        Comparison of model reconstruction efficiencies for full-event (\(\varepsilon_{t\bar{t}}\)), top quark (\(\varepsilon_t\)), and \(W\) boson (\(\varepsilon_W\)) reconstruction.
        The values for SPANet, Topograph, and HyPER are taken from Reference~\cite{hyper}.
        The uncertainties of \pairton were calculated by varying the initial random initialisation and shuffling the training batches five times.
        Since only statistical uncertainties of the test dataset size were considered for the other models, we only quote the central value for these, and note that the statistical uncertainties are generally around \num{0.001}.
        While only fully reconstructible events were considered for the full-event reconstruction efficiencies, reconstructible \(W\) bosons and top quarks from partially reconstructible events were additionally considered for their respective particle reconstruction efficiencies.
        \label{table:event_tpr}
    }
    \begin{tabular}{l l c c c c}
        \toprule
        & Jets  & SPANet & Topograph & HyPER & \pairton \\
        \midrule
        \multirow{4}{*}{\(\varepsilon_{t\bar{t}}\)}
        & 6           & \(0.801\) & \(0.805\) & \(0.816\) & \(\textbf{0.8298} \pm 0.0004\) \\
        & 7           & \(0.654\) & \(0.649\) & \(0.672\) & \(\textbf{0.6988} \pm 0.0011\) \\
        & \(\geq 8\)  & \(0.487\) & \(0.472\) & \(0.513\) & \(\textbf{0.5459} \pm 0.0004\) \\
        & Inclusive   & \(0.651\) & \(0.645\) & \(0.670\) & \(\textbf{0.6944} \pm 0.0003\) \\
        \midrule
        \multirow{4}{*}{\(\varepsilon_t\)}
        & 6           & \(0.667\) & \(0.658\) & \(0.663\) & \(\textbf{0.6760} \pm 0.0008\) \\
        & 7           & \(0.649\) & \(0.638\) & \(0.654\) & \(\textbf{0.6693} \pm 0.0005\) \\
        & \(\geq 8\)  & \(0.586\) & \(0.573\) & \(0.601\) & \(\textbf{0.6195} \pm 0.0002\) \\
        & Inclusive   & \(0.640\) & \(0.629\) & \(0.644\) & \(\textbf{0.6589} \pm 0.0005\) \\
        \midrule
        \multirow{4}{*}{\(\varepsilon_W\)}
        & 6           & \(0.688\) & \(0.683\) & \(0.697\) & \(\textbf{0.7010} \pm 0.0018\) \\
        & 7           & \(0.671\) & \(0.662\) & \(0.683\) & \(\textbf{0.6925} \pm 0.0012\) \\
        & \(\geq 8\)  & \(0.608\) & \(0.594\) & \(0.628\) & \(\textbf{0.6409} \pm 0.0008\) \\
        & Inclusive   & \(0.662\) & \(0.653\) & \(0.675\) & \(\textbf{0.6826} \pm 0.0013\) \\
        \bottomrule
    \end{tabular}
\end{table}

Using these metrics, \pairton significantly outperforms current approaches when compared to SPANet, Topograph, and HyPER as shown in \cref{table:event_tpr}, where the efficiency figures for the reference methods are taken from Reference~\cite{hyper}.
The largest improvements can be observed at high jet multiplicities, where reconstructing the full \(t\bar{t}\) topology is especially difficult due to large jet combinatorics and a crowded detector environment.

Since TIGER~\cite{tiger} aims to be topology agnostic, only the efficiency where the model had been trained on all events is included in Reference~\cite{tiger}.
However, while \pairton requires a complete ground-truth assignment and likewise predicts full assignments, TIGER is trained on the entire dataset and can predict partial reconstructions.
As a result, a fair direct comparison of efficiencies is not possible, since both the training statistics and the models differ between the two.
We note, however, that the efficiencies achieved by TIGER are comparable to those of HyPER.

For \pairton, uncertainties were estimated by retraining the model five times with different random initialisations and shuffled training batches.
The resulting variations are small, indicating stable optimisation and low sensitivity to initialisation.
Equivalent retraining statistics were not available for the comparison methods.
Instead, their uncertainties reported in Reference~\cite{hyper} correspond to statistical uncertainties of the test dataset only and are therefore not included in \cref{table:event_tpr}.

\begin{figure}[tb]
    \centering
    \includegraphics[width=0.45\textwidth]{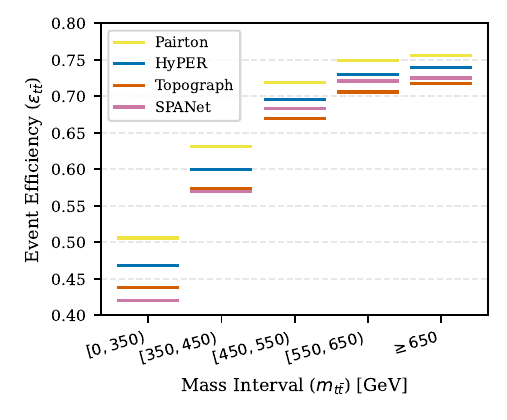}
    \caption{%
        Event reconstruction efficiencies of the different models in dependence on \(m_{t\bar{t}}\), the invariant mass of the \(t\bar{t}\) system as determined from truth-matched final-state reconstructed jets.
        While \pairton consistently outperforms the other methods with HyPER showing the second-highest efficiencies, Topograph outperforms SPANet in the low-mass region while their ordering is reversed for high \(m_{t\bar{t}}\).
        The performance metrics of HyPER, SPANet, and Topograph are taken from Reference~\cite{hyper}.
        \label{fig:tpr_invariant_mass}
    }
\end{figure}

In addition to the phase-space inclusive reconstruction efficiencies, full-event reconstruction efficiencies were also calculated in dependence on the mass of the \(t\bar{t}\) system as determined from the truth-matched reconstructed jets in fully reconstructible events.
This differential efficiency is especially interesting to analyses interested in only measuring parts of the invariant mass spectrum, such as quantum entanglement measurements~\cite{entanglement_ATLAS, entanglement_CMS} or some analyses investigating toponium states at the \(t\bar{t}\) production threshold~\cite{toponium_ATLAS}.
The corresponding comparison of \pairton with existing methods is shown in \cref{fig:tpr_invariant_mass}.
Similar to the comparisons split by jet multiplicity, we find that \pairton outperforms earlier methods across the board, with the largest gains observed for low \(m_{t\bar{t}}\), where the reconstruction efficiencies are lowest.

\subsection*{Ablation Studies}
To evaluate the individual contributions of our architectural choices and the iterative reconstruction procedure, we performed a series of ablation studies, whose full details are provided in \cref{sec:appendix_model_ablation}.
First, we compared the core pairformer architecture against a standard transformer and a bias-augmented transformer baseline.
The pairformer yielded substantial improvements, achieving roughly \qty{5}{\percent} absolute increase in full-event reconstruction efficiency over the standard transformer. 
Second, we tested the impact of our iterative prediction formulation.
We found that our 3-step generative process outperforms both single-step prediction and a hyperedge based prediction modelled after the prediction head used in Reference~\cite{hyper}.
Furthermore, simply increasing the model depth to match the computational cost of the iterative process did not yield any increased performance.
This confirms that the observed state-of-the-art gains originate fundamentally from the iterative nature of our reconstruction algorithm in addition to the performance gains due to the pairformer architecture.

\section{\label{sec:conclusion} Conclusion}

We have presented \pairton, a new approach to the reconstruction of short-lived particles by reformulating the problem as a discrete diffusion process over graphs.
By factorising the joint distribution over reconstructed particles and training a masked edge predictor to learn the corresponding conditional distributions, we obtain a probabilistic formulation for event reconstruction.

Conceptually, \pairton differs in two important ways from previous approaches.
First, inspired by absorbing state discrete diffusion language modelling, we perform reconstruction iteratively by conditioning on previously partially reconstructed structures.
This enables the model to incorporate global consistency constraints in a natural and probabilistically sound manner.
Second, to facilitate this conditioning, our model takes as additional inputs previously predicted structures.
In order to achieve this more smoothly we opt for a prediction method, that purely relies on the partially predicted adjacency matrices, since these can be seamlessly used as pairwise edge features in a graph neural network.

Using a pairformer-based architecture with dynamically updated pairwise representations, we achieve state-of-the-art performance on the fully hadronic \(t\bar{t}\) reconstruction benchmark.
Across all jet multiplicities, \pairton consistently outperforms current methods in full-event, top quark, and \(W\) boson reconstruction efficiencies.
The largest improvements can be observed at high jet multiplicities.

The presented framework is general and can be extended beyond the all-hadronic \(t\bar{t}\) system.
While our masking process follows a relatively straightforward hierarchical strategy, the discrete diffusion formulation enables the definition of arbitrary noising processes provided they transform the fully visible adjacency matrices to fully masked ones.
In particular, more complex topologies, such as \(t\bar{t}H\) reconstruction, would inherently be less linear, since the ordering of predicting \(W\) boson edges and Higgs boson edges could be made interchangeable.

We believe that iterative graph reconstruction provides a powerful and flexible paradigm for particle reconstruction, bridging ideas from modern generative modelling and high-energy physics event interpretation.
\\[0.8\baselineskip]

\section{\label{sec:acknowledgments} Acknowledgments}
We would like to thank Ethan Simpson and Zihan Zhang for stimulating discussions providing insights into \mbox{HyPER} and for their kind support in making available to us additional reference model performance metrics.

This work is supported by the European Union's Horizon Europe research and innovation program under the Marie Sk{\l}odowska-Curie grant agreement No 101168829 called ``Challenging AI with Challenges from Physics: How to solve fundamental problems in Physics by AI and vice versa (AIPHY)''.
We would also like to acknowledge funding through the SNSF project grant 200020\_212127 called ``At the two upgrade frontiers: machine learning and the ITk Pixel detector''.

The training and evaluation of the models developed in this work were performed at the University of Geneva using the Baobab high-performance computing cluster.

\bibliography{bibliography}% Produces the bibliography via BibTeX.

\appendix
\crefalias{section}{appendix}
\section{\label{sec:appendix_proof} Connection between Masked Denoising and Auto-Regressive Factorisation}

In this section, we clarify the relationship between the training objective and the learned conditional distributions.

The training objective is given by
\begin{equation}
    \mathcal{L}(\theta)
    = \EE
    \left[
        \sum_{\alpha} \sum_{i,j \in \mathcal{V}}
        \delta_{\mathrm{M}}(\tilde{a}_{ij}^\alpha)
        \, \ell_{\mathrm{bce}}\!\left(
            a_{ij}^\alpha,\,
            (f_\theta(x, \tilde{A}))_{ij}^\alpha
        \right)
    \right].
\end{equation}

It is well known that, for a Bernoulli variable \(y\), the minimiser of the binary cross-entropy satisfies
\begin{equation}
    f^*(x) = \PP(y = 1 \mid x).
\end{equation}
Applying this result entry-wise, we obtain that the optimal predictor satisfies
\begin{equation}
    (f^*(x, \tilde{A}))_{ij}^\alpha
    = \PP(a_{ij}^\alpha = 1 \mid x, \tilde{A}).
\end{equation}

Thus, the model learns the conditional distribution of each adjacency entry given the observed structure and input features.

The masking process defines a distribution over partially observed adjacency matrices,
\begin{equation}
    q_t(\tilde{A} \mid A),
\end{equation}
which determines which variables are observed and which are predicted.

In our setting, the masking process is structured such that:
\begin{itemize}
    \item when no particles are revealed, all entries are masked,
    \item when one \(W\) boson is revealed, all entries implied by that \(W\) boson are observed,
    \item when both \(W\) bosons are revealed, all entries implied by them are observed.
\end{itemize}

Therefore, the learned conditional distributions take the form
\begin{equation}
    \PP(a_{ij}^\alpha \mid x, \tilde{A}),
\end{equation}
where \(\tilde{A}\) encodes the currently revealed particle assignments.
Because the masking process reveals particles in stages, the model predictions can be associated explicitly to the conditional probabilities:
\begin{itemize}
    \item
        The edges of a \(W\) boson correspond to modelling \mbox{\(\PP\lrp{\lrc{W_1 = \lrp{i,j}} \cup \lrc{W_2 = \lrp{i,j}} \mid x}\)}, and since the events \mbox{\(W_1 = \lrp{i,j}\)} and \mbox{\(W_2 = \lrp{i,j}\)} are mutually exclusive, this becomes \mbox{\(\PP\lrp{W_1 \mid x} + \PP\lrp{W_2 \mid x}\)}.
        Since the ordering of the \(W\) bosons is arbitrary this effectively models \mbox{\(\PP\lrp{W \mid x}\)}.
    \item
        The probabilities of the edges corresponding to the second \(W\) boson map directly to \mbox{\(\PP\lrp{W_2 = \lrp{i,j} \mid W_1, x}\)}.
    \item
        The top quark edges connecting a node \(k\) of a \(b\) quark to an already existing \mbox{\(W = \lrp{i,j}\)} and a second visible \(W\) correspond to \mbox{\(\PP(\lrc{b_1 = k} \mid \lrc{W_1 = \lrp{i,j}},  W_2, x)\)}.
\end{itemize}

Thus, the training procedure is consistent with learning the conditional distributions appearing in an auto-regressive factorisation of the joint distribution over particle assignments. 
By reversing this specific masking chain, the inference procedure exactly mirrors the chain rule factorisation of the joint distribution presented in \cref{eq:factorization}.

\section{\label{sec:appendix_hyperparam} Model Hyperparameters}

\begin{table}[htb]
    \centering
    \caption{%
        Summary of the most important hyperparameters.
        \label{table:h_params}
    }
    \begin{tabular}{lc}
        \toprule
        Name                         & Value               \\
        \midrule
        Number of pairformer blocks  & \num{4}             \\
        Sequence embedding dimension & \num{128}           \\
        Pairwise embedding dimension & \num{64}            \\
        Optimiser                    & AdamW               \\
        Learning rate                & \num{0.001}         \\
        Learning rate scheduler      & Cosine annealing    \\
        Optimiser steps              & 826\,k (100 epochs) \\
        Batch Size                   & \num{256}           \\
        \bottomrule
    \end{tabular}
\end{table}

For the hyperparameters of the pairformer we generally follow the original hyperparameters from Reference~\cite{alphafold3}.
However, we use a much smaller architecture and replace \texttt{LayerNorm} with the now more common \texttt{RMSNorm}.
A summary of hyperparameters used is included in \cref{table:h_params}.
Overall, the parameters chosen resulted in a model with approximately \num{1.2e6} learnable parameters.

All models were implemented in \texttt{python v3.12.3} (\texttt{\href{https://www.python.org}{https://www.python.org}}) using \texttt{pytorch v2.8.0}~\cite{pytorch} and trained using \texttt{pytorch-lightning v2.6.1}~\cite{pytorch_lightning}.
For workflow management we used \texttt{snakemake}~\cite{snakemake}.
Our models were trained on a single RTX~3080 unless otherwise specified.
Despite the iterative nature of our model, it achieves a reconstruction throughput of roughly \num{2000} events per second during inference.

\section{\label{sec:appendix_model_ablation} Model Ablation Studies}
We have performed two ablation studies on details of the neural network architecture and the reconstruction method.
In both studies, efficiencies were evaluated on the validation set in order to not bias the performance estimates provided on the test set.
As a result, the listed performance of our final model differs slightly between the ablation studies and the efficiency comparison listed in \cref{table:event_tpr}, which was evaluated on the test set.

In the first study on the choice of neural network architecture, we compared the transformer architecture, which has been widely adopted in many physics applications, a transformer with a bias term in the attention matrix similar to the ParT architecture used for jet-tagging, and the pairformer architecture.

To keep the compute budgets roughly similar for a fair model comparison, we kept the dimension of the neural networks fixed at \num{128} for the sequence embeddings, while a dimension of \num{64} was used for the pairwise embedding.
To compensate for the fact these architectures do not have the same number of parameters per layer, we decided to increase the number of layers to eight for the transformer (six for the transformer with bias), while the pairformer maintained four layers.
We found that this gave roughly the same throughput for each model during training.
We note that this is not the only way to increase complexity of the alternative models, but a full hyperparameter search for all models was outside the scope of this ablation study.
The remaining hyperparameters were held constant between the models.

Since the standard transformer is unable to use pairwise features, we decided to not make information from the partially masked adjacency matrices available to any model during the ablation study, as opposed to the architecture of the final model described in \cref{sec:method}.
Stated mathematically, for this ablation study we define our models such that
\begin{equation}
    f_\theta(x, \tilde{a}) = f_\theta(x, \emptyset).
\end{equation}
To capture the gain achieved from access to the partially masked adjacency matrices, we also trained an ``Iterative pairformer'', which is conditioned on these matrices.

\begin{table}[tb]
    \centering
    \caption{%
        Comparison of the reconstruction efficiencies from the ablation study comparing different neural network architectures.
        The efficiencies shown are the maximum efficiencies achieved on the validation set during model training.
        \label{table:ablation_model}
    }
    \begin{tabular}{lccc}
        \toprule
        Model                 & \(\varepsilon_{t\bar{t}}\) & \(\varepsilon_{t}\) & \(\varepsilon_W\) \\
        \midrule
        Transformer           & \num{0.635}                & \num{0.623}         & \num{0.655}       \\
        Transformer with bias & \num{0.668}                & \num{0.649}         & \num{0.678}       \\
        Pairformer            & \num{0.688}                & \num{0.656}         & \num{0.682}       \\
        Iterative pairformer  & \num{0.692}                & \num{0.661}         & \num{0.684}       \\
        \bottomrule
    \end{tabular}
\end{table}

\cref{table:ablation_model} shows convincingly that the pairformer architecture achieves superior performance compared to the more traditional transformer-based architectures.
The conditioning of the iterative pairformer leads to a further performance improvement.

\begin{figure*}[tb]
    \centering
    \includegraphics[width=\textwidth]{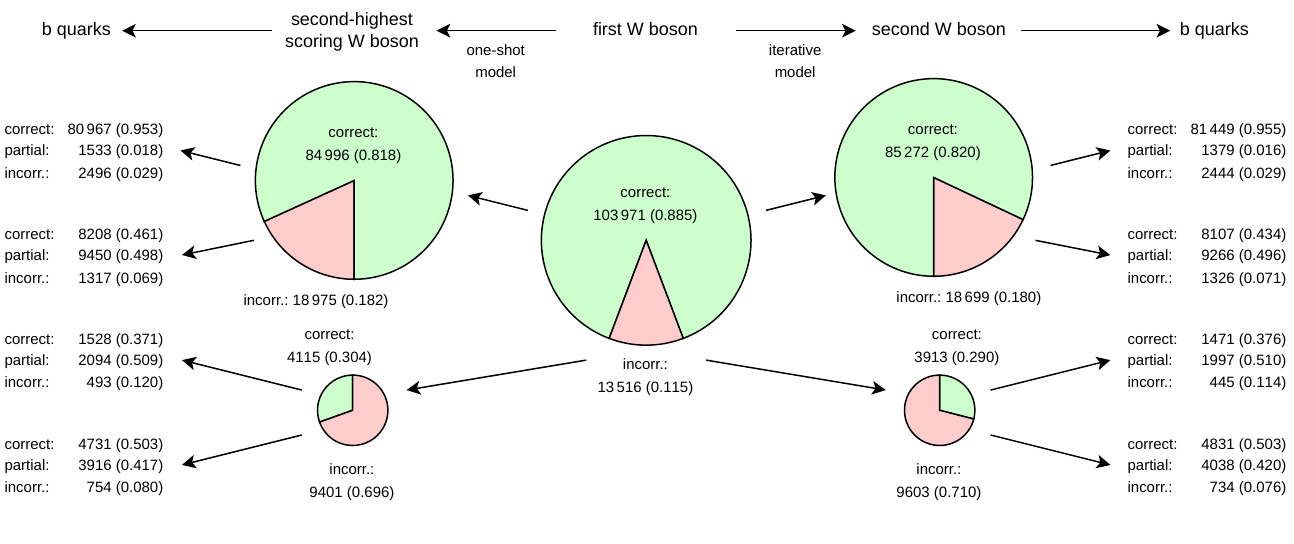}
    \vspace*{-6ex}
    \caption{%
        Comparison of the reconstruction performance of the 3-step iterative pairformer model (towards the right) and a one-shot pairformer model (towards the left) on a total of \num{117487} fully reconstructible all-hadronic \(t\bar{t}\) events of the validation set.
        To directly compare the model behaviour, the reconstruction of the first \(W\) boson is common between the two methods.
        Afterwards, the iterative model reconstructs further particles in additional model passes using partial adjacency matrices from previous iterations as described in \cref{sec:method}.
        Meanwhile, all particles are reconstructed from the first model pass in a greedy approach for the one-shot model, whereby the first \(W\) boson selected is the one maximising the edge assignment heuristic detailed in \cref{sec:appendix_algorithm}.
        In each reconstruction step, the number of correct and incorrect assignments, as well as their proportion with respect to the previous assignment outcome, are shown.
        For the \(b\) quark assignment, the number and proportion of partial correct assignments are additionally shown while respecting consistency in the top quark matching between possible correctly reconstructed \(W\) bosons and \(b\) quarks.
        \label{fig:oneshot-iterative-reco-comp}
    }
\end{figure*}

This observation is also exemplified in our second ablation study, which specifically compares the effects of our iterative prediction method against other reconstruction methods.
As a baseline, we included the unconditioned pairformer from the previous ablation study and added the iterative pairformer running on three steps:
the reconstruction of the first \(W\) boson, the reconstruction of the second \(W\) boson, and the simultaneous reconstruction of the two \(b\) quarks.
In addition to this 3-step iterative model, we included a 4-step generative procedure where the \(b\) quark prediction was factorised instead of the simultaneous prediction of both \(b\) quarks.

We also compare to the hyperedge-based reconstruction method utilised by HyPER, where we use a pairformer architecture as the embedding graph neural network instead of the graph neural network trained in the original HyPER paper.
Since the HyPER method needs to materialise the hyperedges, we found that the resulting model required slightly more memory than our reconstruction method, which meant it just barely did not fit on an RTX~3080 GPU.
As a consequence, it was trained on an RTX~3090 instead.

In order to investigate the effect of a larger test-time compute budget, we included a deep pairformer architecture as a last reference model.
This model consists of twelve pairformer blocks as opposed to the four blocks used in the baseline model.
Similar to the model with the hyperedge-based reconstruction head, this larger pairformer architecture was trained on an RTX~3090 using the same batch size as the other models.

\begin{table}[htb]
    \centering
    \caption{%
        Comparison of the reconstruction efficiencies from the ablation study comparing different reconstruction methods.
        The efficiencies shown are the maximum efficiencies achieved on the validation set during model training.
        \label{table:ablation_diffusion} 
    }
    \begin{tabular}{lccc}
        \toprule
        Model                       & $\varepsilon_{t\bar{t}}$ & $\varepsilon_{t}$ & $\varepsilon_W$ \\
        \midrule
        Pairformer                  & $0.688$                  & $0.656$           & $0.682$         \\
        Pairformer HyPER            & $0.688$                  & $0.657$           & $0.681$         \\
        Iterative pairformer 3-step & $0.692$                  & $0.661$           & $0.684$         \\
        Iterative pairformer 4-step & $0.691$                  & $0.660$           & $0.684$         \\
        Pairformer deep             & $0.687$                  & $0.658$           & $0.684$         \\
        \bottomrule
    \end{tabular}
\end{table}

The iterative predictive model outperforms the non-iterative baseline and the hyperedge-based reconstruction, as shown in \cref{table:ablation_diffusion}.
Additionally, the iterative models also outperform the deeper alternative, showcasing that our performance gain cannot exclusively be attributed to a higher test-time compute budget.
From a step-by-step comparison of the reconstruction performance between jointly trained one-shot pairformer and 3-step iterative models shown in \cref{fig:oneshot-iterative-reco-comp}, it can be concluded that the improvement of the iterative model mainly stems from a better prediction performance after successful previous predictions.
At the same time, the iterative model shows a poorer ability to recover from incorrect assignments, which may be due to the current training procedure, in which the iterative model only trains on correct partial adjacency matrices for the assignments of the second \(W\) boson and the \(b\) quarks.
Augmenting the training to make use of both correct and incorrect previous assignments might therefore lead to further performance improvements.

Since we observe similar performance between the 3-step and the 4-step models, we opted for a 3-step iterative reconstruction in the final model to increase event throughput.

\section{\label{sec:appendix_algorithm} Algorithm for Graph Reconstruction}

Here we provide pseudo-code for how we reconstruct edges given the model outputs.
The decoding step uses a greedy approximation, motivated by the optimal assignment in the case of independent predictions which becomes a tractable problem.
This means that we select the edge which locally maximises
\begin{equation}\label{eq:post_processing_obj}
    \prod_{i\leq j} p_{i,j}^{Z_{i,j}} \lrp{1-p_{i,j}}^{1-Z_{i,j}} \propto \sum_{i \leq j} Z_{i,j} \logit p_{i,j}
\end{equation}
in a greedy manner at each prediction step in the notation of \cref{eq:logit_heuristic}.
We note that other reconstruction algorithms could be explored, and that attempting to maximise the above quantity might not yield the best reconstruction algorithm in general.

\begin{algorithm}[tb]
    \DontPrintSemicolon
    \caption{%
        Algorithm for choosing \(W\) boson assignments from \(W\) adjacency logits.
        \label{alg:choose_w}
    }
    \KwIn{\(\sigma^W_{ij}\): Logits for the \(W\) adjacency matrix}
    \KwIn{\((W^\ast_1, W^\ast_2)\): Optional \(W\) boson indices}
    \(g_{ij}^W \gets \sigma^W_{ij} - \sigma^W_{ii} - \sigma^W_{jj}\)\;
    \(g_{ii}^W \gets -\infty\)\tcp*{W boson assignments only}
    \If{\(W^\ast\) exists}{
        \(g_{W^\ast_1\,k}, g_{k\,W^\ast_1} \gets -\infty \quad \forall k\)\tcp*{Force new W boson}
        \(g_{W^\ast_2\,k}, g_{k\,W^\ast_2} \gets -\infty \quad \forall k\)\;
    }
    \(W = \left( W_1, W_2 \right) \gets \arg \max_{i,j} g_{ij}^W\)\;
    \Return \( W \)
\end{algorithm}

\begin{algorithm}[tb]
    \DontPrintSemicolon
    \caption{%
        Algorithm for choosing \(b\) quark assignments from top logits and \(W\) boson assignments.
        \label{alg:choose_b}
    }
    \KwIn{\(\sigma^t_{ij}\): Logits for top adjacency matrix}
    \KwIn{\(W^{(1)}, W^{(2)}\): Pairs of chosen \(W\) boson jets}
    \(g_{k}^{b(1)} \gets \sigma^t_{k\,W^{(1)}_1} + \sigma^t_{k\,W^{(1)}_2} - \sigma^t_{kk}\)\;
    \(g_{l}^{b(2)} \gets \sigma^t_{l\,W^{(2)}_1} + \sigma^t_{l\,W^{(2)}_2} - \sigma^t_{ll}\)\;
    \(c_{kl} \gets g_{k}^{b(1)} + g_{l}^{b(2)}\)\;
    \(c_{kk} \gets -\infty \quad \forall k\)\tcp*{Choose different b quarks}
    \(( b^{(1)}, b^{(2)}) \gets \arg \max_{k,l} c_{kl}\)\;
    \Return \(b^{(1)}, b^{(2)}\)
\end{algorithm}

For the \(W\) boson assignment we simply select the edge indices \(\hat{i}, \hat{j}\) which locally maximise 
\begin{equation}
    \hat{i}, \hat{j} = \arg \max_{i, j} \sigma_{i,j} - \sigma_{i,i} - \sigma_{j,j}.
\end{equation}
Therefore, the assignment takes into account both the possible edge between the two nodes \(i\) and \(j\) as well as the possible self-edges of each node.
The exact procedure is described in \cref{alg:choose_w}.

\begin{algorithm}[h]
    \DontPrintSemicolon
    \caption{%
        Overall heuristic algorithm for choosing top quark and \(W\) boson assignments.
        \label{alg:inference}
    }
    \KwIn{\(f_\theta^{W/t}\): Model to predict logits from data and partial adjacency matrices}
    \(\sigma^W \gets f^W_\theta(x, \emptyset)\)\;
    \(W^{(1)} \gets\) \FChooseW{$\sigma^W$}\;
    \(\sigma^W \gets f^W_\theta\lrp{x, \tilde{A}(W^{(1)})}\)\tcp*{Update W logits}
    \(W^{(2)} \gets\) \FChooseW{$\sigma^W, W^{(1)}$}\;
    \(\sigma^t \gets f^t_\theta\lrp{x, \tilde{A}(W^{(1)}, W^{(2)})}\)\tcp*{Compute t logits}
    \(b^{(1)}, b^{(2)} \gets\) \FChooseB{$\sigma^t, W^{(1)}, W^{(2)}$}\;
    \Return \(W^{(1)}, W^{(2)}, b^{(1)}, b^{(2)}\)
\end{algorithm}

For selecting the \(b\) quarks for the already chosen \(W\) boson, we again aim to maximise \cref{eq:post_processing_obj}.
However, since we already have an edge, we will instead add the contributions from the two edges connecting to the new node, while subtracting the self-contribution of the node.
We do this for both \(b\) quarks and select the pair of \(b\) quarks which maximises the sum of their contributions.
The pseudo-code for this is shown in \cref{alg:choose_b}.

\FloatBarrier
For the full reconstruction algorithm we interchangeably predict the logits of the edges and then apply the appropriate selection algorithm until we have selected all particles as shown in \cref{alg:inference}.

\end{document}